\documentclass[12pt,preprint]{aastex}

\begin{document}
\newcommand{\lens}{PKS~1830--211}
\def\gtorder{\mathrel{\raise.3ex\hbox{$>$}\mkern-14mu
             \lower0.6ex\hbox{$\sim$}}}

\title{PKS~1830--211: A Face-On Spiral Galaxy Lens}

\author{
Joshua N.\ Winn\altaffilmark{1},
Christopher S.\ Kochanek\altaffilmark{1},
Brian A.\ McLeod\altaffilmark{1},
Emilio E.\ Falco\altaffilmark{2},
Christopher D.\ Impey\altaffilmark{3},
Hans-Walter Rix\altaffilmark{4}
}

\altaffiltext{1}{Harvard-Smithsonian Center for Astrophysics, 60
Garden St., Cambridge, MA 02138; {\tt jwinn, kochanek,
bmcleod@cfa.harvard.edu}}

\altaffiltext{2}{Smithsonian Institution, Whipple Observatory, 670 Mt
Hopkins Road, P.O.\ Box 97, Amado, AZ 85645; {\tt
falco@cfa.harvard.edu}}

\altaffiltext{3}{Steward Observatory, University of Arizona, Tucson,
AZ 85721; {\tt cimpey@as.arizona.edu}}

\altaffiltext{4}{Max-Planck-Institut f\"{u}r Astronomie,
K\"{o}nigsstuhl 17, D-69117 Heidelberg, Germany; {\tt
rix@mpia-hd.mpg.de}}

\begin{abstract}

We present new Hubble Space Telescope images of the gravitational lens
\lens, which allow us to characterize the lens galaxy and update the
determination of the Hubble constant ($H_0$) from this system.  The
$I$-band image shows that the lens galaxy is a face-on spiral galaxy
with clearly delineated spiral arms.  The southwestern image of the
background quasar passes through one of the spiral arms, explaining
the previous detections of large quantities of molecular gas and dust
in front of this image.  The lens galaxy photometry is consistent with
the Tully-Fisher relation, suggesting the lens galaxy is a typical
spiral galaxy for its redshift.  The lens galaxy position, which was
the main source of uncertainty in previous attempts to determine
$H_0$, is now known precisely.  Given the current time delay
measurement and assuming the lens galaxy has an isothermal mass
distribution, we compute $H_0=44\pm 9$~km~s$^{-1}$~Mpc$^{-1}$ for an
$\Omega_m =0.3$ flat cosmological model.  We describe some possible
systematic errors and how to reduce them.  We also discuss the
possibility raised by Courbin et al.\ (2002), that what we have
identified as a single lens galaxy is actually a foreground star and
two separate galaxies.

\end{abstract}


\section{Introduction}

Although the Hubble Space Telescope (HST) Key Project to Measure the
Hubble Constant has been successfully completed \citep{freedman01}, it
is important to pursue completely different methods of determining the
Hubble constant ($H_0$).  Independent methods provide a consistency
check, and may eventually surpass the 10\% accuracy of the local
distance-scale methods employed by the Key Project.  It is also
important to measure the expansion rate directly at cosmological
redshifts, in case the Galaxy is in a locally underdense or overdense
region \cite[see, e.g.,][]{wu95}.  These goals will grow in importance
in the coming years due to the degeneracies in analyses of cosmic
microwave background anisotropies between many cosmological parameters
and $H_0$ \cite[e.g.,][]{bond94,eisenstein98}.

A promising approach to determining $H_0$ that is independent of the
local distance scale uses gravitational lens time delays
(\citealt{refsdal64}; for recent summaries see, e.g.,
\citealt{schechter00}, \citealt{koopmans99}).  This method is
ultimately limited by systematic uncertainties in the mass models of
the lens galaxies, but a necessary first step is to obtain the basic
observational constraints---astrometry, photometry, and redshifts of
the lens and source---for the systems with measured time delays.

For the gravitational lens \lens, this has been especially challenging
because the system is near the Galactic plane ($b=-5\fdg 7$).  As a
result, most of the information has come from radio and infrared
observations.  \citet{rao88} suggested it was a lens due to its radio
morphology. \citet{subrahmanyan90} and \citet{jauncey91} provided more
conclusive evidence; the system has two bright point sources (NE and
SW) embedded in a fainter Einstein ring.  Strong molecular absorption
features were detected at $z_l=0.886$ \citep{wiklind96,gerin97},
largely in front of SW \citep{frye97,swift01}, and presumably due to
the lens galaxy.  \citet{lovell96} found $z_a=0.19$ \ion{H}{1}
absorption in front of NE, of unknown provenance.  \citet{lovell98}
measured a time delay of $26^{+4}_{-5}$~days between NE and SW at
8.6~GHz.  By deconvolving ground-based infrared images,
\citet{courbin98} detected the quasar images and found hints of the
lens galaxy.  \citet{lidman99} determined the quasar redshift of
$z_s=2.507$ by infrared spectroscopy.  From the presence and structure
of the molecular absorption system, \citet{wiklind98} argued that the
lens galaxy is probably a spiral galaxy seen nearly face-on, which is
consistent with the large differential extinction between NE and SW
($\Delta E_{B-V} \approx 3$) observed by \citet{falco99}.  However,
although \citet{lehar00} detected the lens galaxy in $H$-band and
shallow $I$-band HST images, they could not determine its position or
structure accurately.  These authors found that the uncertainty in the
lens galaxy position dominated the uncertainty in the value of $H_0$
inferred from the time delay.

In this paper we present new HST optical images that confirm the lens
galaxy is a face-on spiral galaxy, and allow its position and optical
magnitudes to be measured accurately.  These data are discussed in
\S~\ref{sec:lensgalaxy}.  In \S~\ref{sec:tullyfisher}, we place the
lens galaxy on the Tully-Fisher relation using the HST photometry and
two different estimators of the galaxy mass---one from the lens
geometry, and one from the velocity shift measured by
\citet{wiklind98}.  In \S~\ref{sec:models}, we incorporate our
measurement of the position of the lens galaxy into a simple lens
model to arrive at an updated estimate of $H_0$.  We also discuss
systematic errors and compare our model with previous models.

After this work was completed, we learned that \citet{courbin02}
independently analyzed the same HST data, along with new and archival
near-IR data.  They argue that what we have identified as the bulge of
the lens galaxy is actually a foreground star, and that there is a
second lens galaxy.  In the final section of this paper, we describe
the strengths, weaknesses, and future tests of these competing
hypotheses, and discuss future prospects for reducing systematic
errors in $H_0$ from this particular gravitational lens.

\section{Observations and data reduction}
\label{sec:lensgalaxy}

\lens~was observed with HST/WFPC2\footnote{ Data from the NASA/ESA
Hubble Space Telescope (HST) were obtained from the Space Telescope
Science Institute, which is operated by AURA, Inc., under NASA
contract NAS~5-26555.} on 2000~September~25 with the F814W filter,
using 4 dithered exposures, for a total integration time of 4800~s.
On 2001~July~11 it was observed with the F555W filter, using 3
dithered exposures, for a total integration time of 2000~s.  In both
cases \lens~was centered on the PC chip.  We cross-registered and
combined the exposures, and rejected cosmic rays, using the Drizzle
algorithm implemented in IRAF \citep{fruchter02}.

\begin{figure}
\figurenum{1}
\plotone{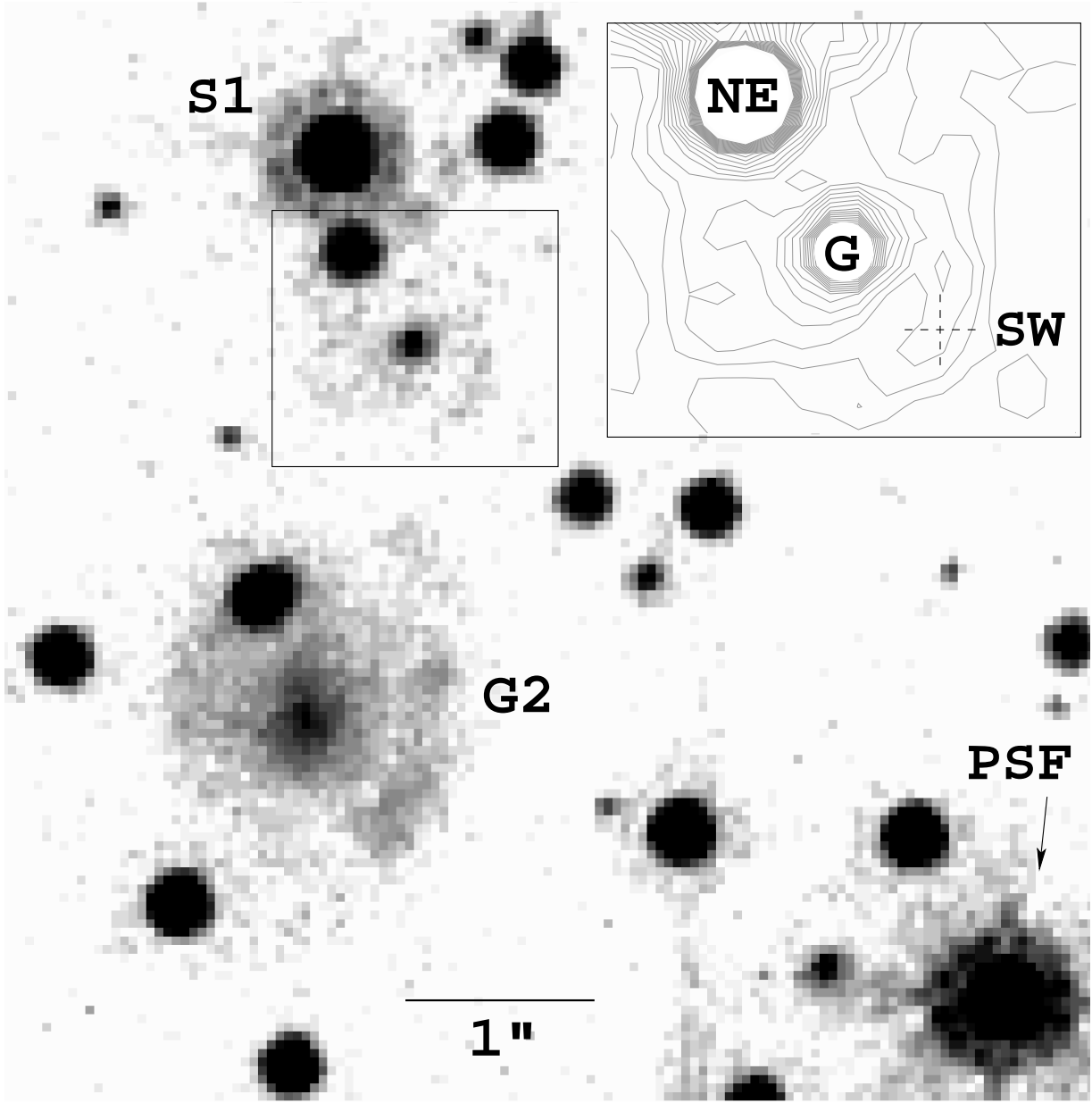}
\caption{ HST/WFPC2 image, filter F814W~$\approx I$. North is up and
east is left.  Note that $1\arcsec$ corresponds to $5.4 h^{-1}$~kpc at
the lens redshift (in a flat $\Omega_m = 0.3$ cosmology).  Inset in
the upper right of the image is a contour representation of the region
surrounding G, based on the same data.  The position of SW, as
determined from radio images, is marked.  }
\label{fig:iband}
\end{figure}

\begin{figure}
\figurenum{2}
\plotone{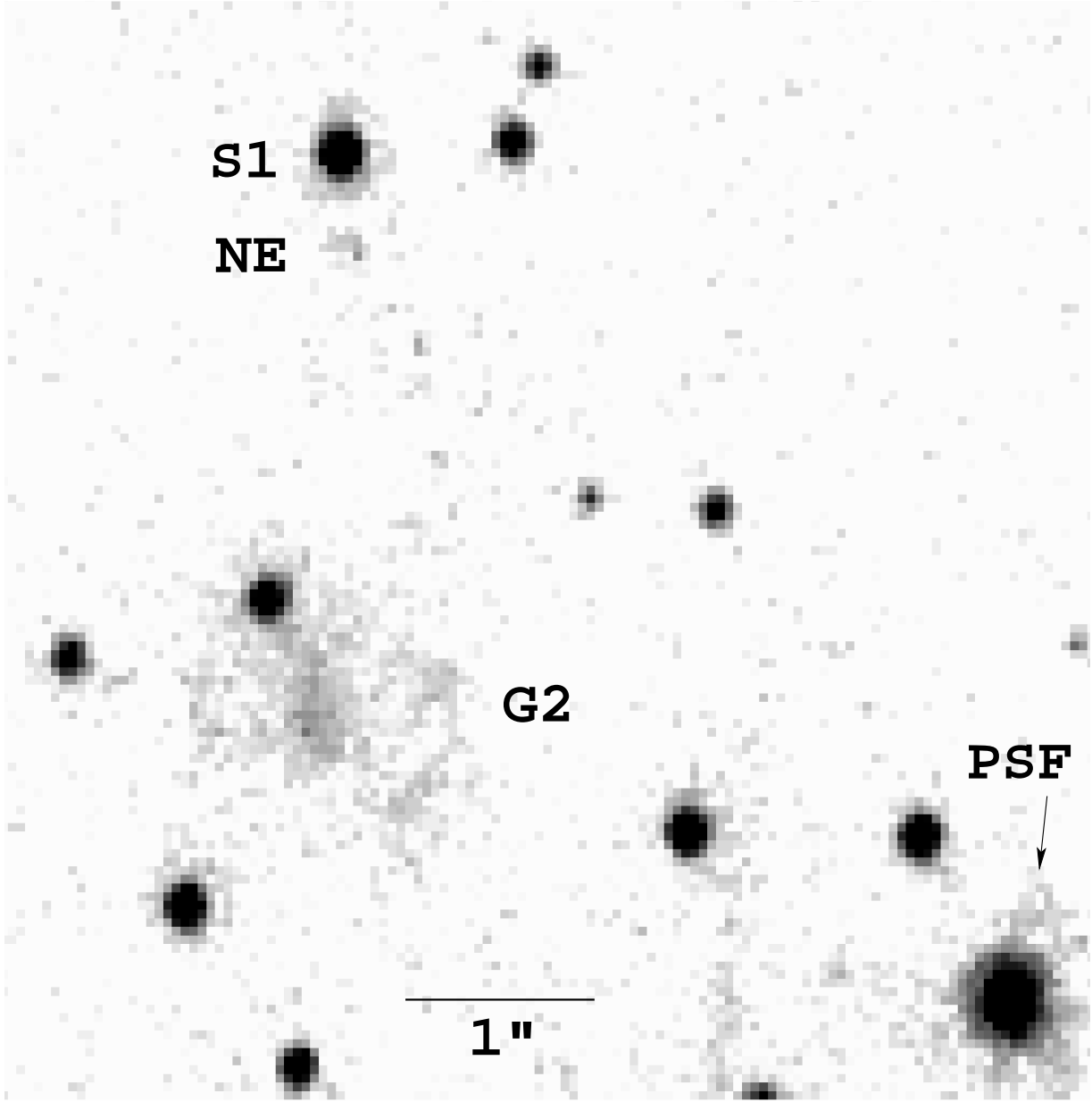}
\caption{ HST/WFPC2 image, filter F555W~$\approx V$. North is up and
east is left. Note that $1\arcsec$ corresponds to $5.4 h^{-1}$~kpc at
the lens redshift (in a flat $\Omega_m = 0.3$ cosmology). }
\label{fig:vband}
\end{figure}

Figure~\ref{fig:iband} shows the F814W (hereafter ``$I$-band'') image.
The lens galaxy, labeled G, is a nearly face-on spiral galaxy
(probably an Sb or Sc).  The bulge of the galaxy is detected with high
significance and is compact.  The position of quasar SW (as determined
from radio images) is covered by one of the spiral arms.  Object S1 is
an M star identified by \citet{djorgovski92}, and G2 is a galaxy
identified by \citet{lehar00}.  Inset in the upper right corner of the
image is a contour map of G, based on the same data, that shows the
western spiral arm more clearly than the grayscale image.
Figure~\ref{fig:vband} shows the F555W (hereafter ``$V$-band'') image.
In this image the bulge of G is only marginally detected.

To measure the position of G relative to NE, which is the crucial
quantity for modeling purposes, we used software written for the
Center for Astrophysics--Arizona Space Telescope Lens Survey (CASTLES;
see, e.g., \citealt{lehar00}) to fit a parameterized model to the
$I$-band image.  A well-exposed star $3\farcs43$ west and $3\farcs94$
south of NE was used as an empirical PSF.  (Although this star might
appear overexposed in Figures~\ref{fig:iband} and \ref{fig:vband},
this is due only to the chosen scaling of gray levels.)

Our model consisted of point sources representing: S1, NE, the central
bulge of G, and all stars within $5\arcsec$ of NE.  We estimate the
uncertainty in each coordinate of the G--NE separation to be
4~milliarcseconds, which is the variance of the separations obtained
by individually fitting the four $I$-band exposures.  The result did
not change significantly when the empirical PSF was replaced by a
theoretical ``Tiny Tim'' PSF \citep{krist97}, nor when G was
represented as a Gaussian or de Vaucouleurs profile instead of a point
source.  The results are given in Table~\ref{tbl:data}, which also
summarizes the basic data for \lens~drawn from this work and the
literature.

We applied the same model to the $V$-band image, but fixed the
relative positions at the $I$-band values and allowed only the fluxes
to vary.  The fluxes of the point sources were converted to magnitudes
using the zero points and CTE correction formulae of
\citet{dolphin00}\footnote{As updated on the web site of A.\ Dolphin,
${\tt http://www.noao.edu/staff/dolphin/wfpc2\_calib/}$, referenced
2001 November 15.}.  The results are given in Table~\ref{tbl:mags}.
The quoted error is the quadrature sum of the variance obtained by
fitting the separate exposures, and 0.04 due to PSF subtraction.  They
do not include an overall uncertainty of $\sim 0.05$ due to zero
points and CTE correction.

Because neither G nor G2 is well described by an analytic profile, we
measured the magnitude of each galaxy within a synthetic circular
aperture, using an image with all the point sources subtracted (except
the point source representing the bulge of G).  In particular, two
point sources were subtracted on the northeast edge of G2, to
represent what appears to be a pair of blended stars.  For G, the
aperture radius was $0\farcs7$; for G2, it was $1\arcsec$.  No CTE
correction was made because the correction procedure for extended
sources is unknown and the correction is probably small
\citep{riess00}.  The errors quoted in Table~\ref{tbl:mags} are due to
uncertainty in the sky level and contributions from the residuals of
the neighboring point sources.

\section{The lens galaxy luminosity and the Tully-Fisher relation}
\label{sec:tullyfisher}

The Tully-Fisher relation is an empirical correlation between the
luminosity and mass of disk galaxies that has been observed for both
nearby galaxies \cite[see, e.g.,][]{tully77,sakai00} and for galaxies
at significant redshifts \cite[e.g.,][]{vogt96,vogt97,ziegler01}.
Typically, the measure of luminosity is rest-frame $M_B$ and the
measure of mass is the circular rotation velocity.  For most galaxies
at $z\sim 1$ the latter measurement is a challenging spectroscopic
project, especially for a nearly face-on spiral galaxy.  Lens galaxies
are special because an accurate mass measurement is available from the
geometry of the background images.  Given the HST photometry and image
configuration of \lens, we can attempt to place the lens galaxy on the
Tully-Fisher relation.

We corrected the $I$-band magnitude of the lens galaxy for Galactic
extinction using the $E_{B-V}$ value of \citet{schlegel98} and
assuming a $R_V=3.1$ extinction law, obtaining $A_I=0.81$.  For the
lens redshift $z_l=0.886$, the $I$-band is nearly centered on the
rest-frame $B$-band, so there is little uncertainty in the
$k$-correction: $B_{\hbox{rest}}$--$I \approx 1.05\pm0.05$~mag.  For a
flat $\Omega_m=0.3$ cosmology with $H_0=65$~km~s$^{-1}$~Mpc$^{-1}$,
the implied rest-frame absolute magnitude of the lens galaxy is
$M_B=-21.7 \pm 0.3$~mag, with the quoted uncertainty entirely due to
the measurement error.  We attempt no extrapolation from our aperture
magnitude to a total magnitude, nor do we correct for internal
extinction, so the resulting luminosity should be an understimate.

The galaxy's circular velocity can be estimated from the image
separation of $0\farcs971$.  For an intrinsically spherical dark
matter halo with a flat rotation curve (a singular isothermal sphere,
or SIS), the image separation is given by $\Delta\theta= 4\pi
(v_c/c)^2 D_{ls}/D_{os}$, where $v_c$ is the circular velocity, and
$D_{ls}$ and $D_{os}$ are (respectively) the lens--source and
observer--source angular diameter distances.  For the SIS model, the
measured image separation corresponds to $v_c=264$~km~s$^{-1}$.  For a
thin Mestel disk, which also has a flat rotation curve, $\Delta\theta=
8 (v_c/c)^2 D_{ls}/D_{os}$ \citep{keeton_kochanek98}, giving
$v_c=331$~km~s$^{-1}$.  The truth is probably between these two
extremes.

With these figures, we find that the lens galaxy is compatible with
the local Tully-Fisher relation as determined by \citet{sakai00}.  For
the SIS model the lens galaxy is $0.1\pm0.3$~mag brighter than
predicted by the T-F relation, and for the Mestel model it is
$0.7\pm0.3$ mag fainter.  The dispersion in the \citet{sakai00} T-F
relation is 0.43~mag.  These results indicate the lens galaxy appears
to be a fairly normal spiral galaxy for its redshift.

An independent estimate of the lens galaxy's circular velocity is
available from the molecular absorption system detected by
\citet{wiklind98}.  They found absorption lines at $z_l=0.886$ in
front of SW, and also a weaker absorption line in front of NE shifted
in rest-frame velocity by $\Delta v = -147$~km~s$^{-1}$.  As they
explained, the observed $\Delta v$ can be used to estimate the
circular velocity $v_c$, using the kinematic relation
\begin{equation}
\Delta v = v_c (\cos\theta_1 - \cos\theta_2) \sin i,
\end{equation}
where $i$ is the inclination, and $\theta_k$ is the angle between the
line of sight $k$ (either NE or SW) and the line of nodes (as measured
in the plane of the galaxy).  The angles $\theta_k$ are related to the
corresponding sky-plane angles $\phi_k$ by
\begin{equation}
\tan(\phi_k) = \tan(\theta_k) \sec i.
\end{equation}

Using these relations, \citet{wiklind98} used the data and models
existing at the time to postulate that the lens galaxy is seen at low
inclination ($i<20\arcdeg$) and is quite massive, and the quasar is at
high redshift ($z>3$).  These postulates have been proven broadly
correct by our data and by the measured quasar redshift ($z=2.507$).

We can now update their determination of $v_c$ using the new data.  We
must borrow a few results from the lens models that will be described
later in \S~\ref{subsec:update}, namely, the inclination of the lens
galaxy ($i=25\arcdeg$) and the orientation of the line of nodes
($86\arcdeg \pm 3\arcdeg$).  From these values and the measured galaxy
position, we derive $\phi_{\mathrm{NE}} = 52\arcdeg$ and
$\phi_{\mathrm{SW}} = 146\arcdeg$.  The resulting circular velocity is
$v_c=255$~km~s$^{-1}$.  The corresponding Tully-Fisher magnitude is
$M_B=-21.5$, which agrees well with the value inferred from the HST
photometry ($-21.7\pm 0.3$).

\section{The lens galaxy position and the Hubble constant}
\label{sec:models}

\subsection{Updated lens models}
\label{subsec:update}

In the most recent modeling effort for this system, \citet{lehar00}
demonstrated a strong correlation between the position of the lens
galaxy, which was poorly known, and the inferred value of $H_0$.  Our
measurement of the position of G allows us to determine $H_0$ with
much higher precision.  At this stage we consider simple models
constrained only by the positions of NE, SW, and G, and the
magnification ratio (and not by the faint Einstein ring, which we
leave for a future undertaking).

We modeled the lens galaxy as a singular isothermal ellipsoid using
software written by \citet{keeton01a}.  With as many parameters as
constraints, the fits were exact.  We estimated the error in each
parameter to be the spread in the results obtained by allowing the
position of G to vary through the full range of its quoted
uncertainty.  The results are given in Table~\ref{tbl:model}.

Using this model and the time delay measured by \citet{lovell98}, we
obtain $h=0.44\pm 0.09$ (where $H_0=100h$~km~s$^{-1}$~Mpc$^{-1}$),
with most of the quoted uncertainty due to the 20\% fractional
uncertainty in the time delay.  This value was computed assuming
$\Omega_m=0.3$ and $\Omega_\Lambda=0.7$, but the dependence on the
cosmological model is weak; for $\Omega_\Lambda=0$ the result is
$h=0.46$.  Either value is significantly lower than the Key Project
value of $0.72\pm 0.08$ \citep{freedman01}, and also lower than
several previous estimates of $H_0$ from this system.  In the rest of
this section we discuss possible systematic modeling errors; in the
next section we compare our model with previous work.

Our simple model might misrepresent the true mass distribution, in
several ways.  First, if there is an unmodeled source of convergence
$\kappa$ (due to, say, a nearby group of galaxies), then the inferred
$h$ is systematically low by the factor $1-\kappa$ \citep[see,
e.g.,][]{gorenstein88}.  \citet{lehar00} identified 4 galaxies besides
the lens galaxy within $20\arcsec$ of \lens, which would contribute a
convergence as high as $\kappa=0.16$ if they all lie at the lens
redshift $z_l=0.886$---although, in the following section, we argue
that the nearest of these galaxies is probably at $z_a=0.19$ where it
has a much smaller effect ($\kappa\sim 0.03$).

\begin{figure}
\figurenum{3}
\plotone{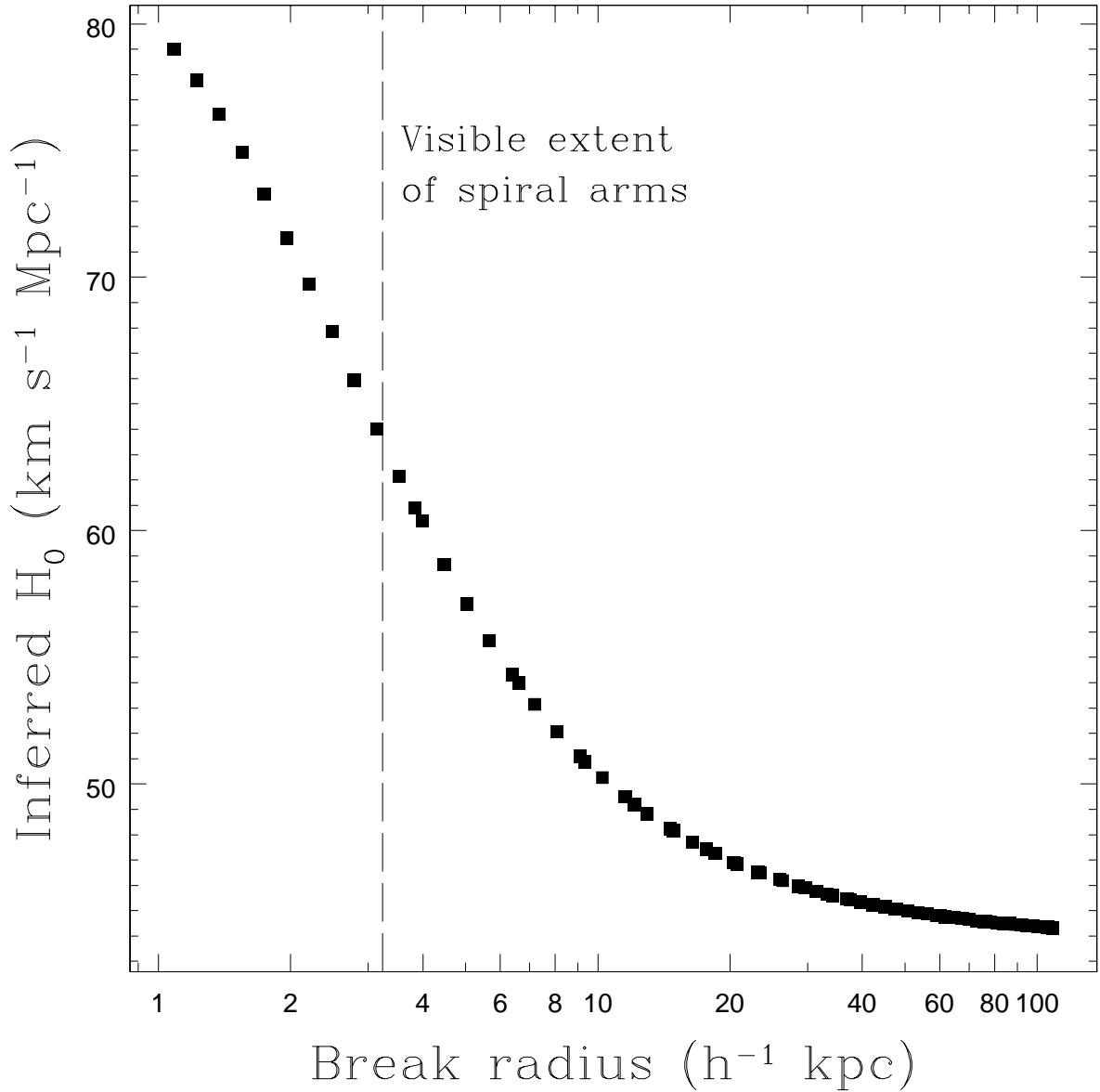}
\caption{ Variation of the inferred $H_0$ with the break radius $a$ in
the pseudo-Jaffe model.  The visible extent of the spiral arms is also
marked.  The horizontal scale has been set by the conversion factor
$5.4 h^{-1}$~kpc per arcsecond, based on a flat $\Omega_m = 0.3$
cosmology.}
\label{fig:h_vs_a}
\end{figure}

Second, if the radial mass distribution is not isothermal, the
inferred $h$ is biased.  If the lens potential $\phi \propto
r^{\beta}$ then the true $h$ is $(2-\beta)$ times the isothermal
estimate of $h$ \citep{witt00}.  A similar bias results if the
isothermal profile is truncated outside a radius comparable to, or
smaller than, the Einstein ring radius.  As a demonstration we modeled
the lens galaxy as a pseudo-Jaffe ellipsoid ($\rho \propto
r^{-2}(r^2+a^2)^{-2}$; see, e.g., Keeton 2001a) with a break radius
$a$.  For $r\ll a$, the pseudo-Jaffe profile has a flat rotation
curve, but for $r\gg a$ it has a Keplerian fall-off.  We fixed $a$ and
allowed all other parameters to vary; the resulting $h$ rises
significantly for $a<b$, where $b$ is the Einstein ring radius.
Figure~\ref{fig:h_vs_a} shows the dependence of the inferred Hubble
constant on the break radius.

However, in order to match the Key Project value of $H_0$, both of the
preceding possibilities require unnatural parameters for the lens
galaxy, which we have established is an apparently normal, massive,
spiral galaxy.  To obtain $h=0.72$ would require $\beta=0.36$; or,
using the pseudo-Jaffe model, $a/b=0.38$, corresponding to $a =
0\farcs42 = 2.3h^{-1}$~kiloparsecs, at the lens redshift (assuming a
flat $\Omega_m = 0.3$ cosmology).  Nearby massive spiral galaxies have
flat rotation curves out to well beyond the visible extent of the
spiral arms \cite[see, e.g.,][]{rubin78,sofue01}.  In the $I$-band
image, the spiral arms of the lens galaxy are detected out to a radius
of at least $0\farcs6 = 3.3h^{-1}$~kpc.

Third, our simple models may misrepresent the angular structure of the
mass distribution, in addition to the radial structure.  To
investigate this possibility we computed a 2-d grid of models
consisting of an SIE embedded in an external shear field.  The
ellipticity of the SIE was stepped from 0.0 to 0.2 (a plausible
maximum value, given the nearly face-on morphology), and its position
angle was stepped from $0\arcdeg$ to $180\arcdeg$, and in each case
the corresponding parameters of the external shear field were
determined.  The resulting range in $h$ was 0.33--0.55.  The range in
$h$ widens, but the inferred $h$ is not systematically large or small.

Fourth, a number of authors have recently argued that some lens
galaxies have small-scale mass substructure (on scales $<10^9 M_\sun$)
that smooth parameterized lens models fail to describe \cite[see,
e.g.,][]{dalal01,keeton01b,bradac01,metcalf01}.  The evidence lies in
the discrepancy between measured and model-predicted flux ratios,
which depend on the local curvature of the lens potential and are
therefore sensitive to perturbations on small angular scales.  This
raises the possibility that the magnification ratio assumed for
\lens~may be grossly in error.  However, the time delay does not
depend strongly on the assumed magnification ratio.  In fact, for
isothermal models, \citet{witt00} showed the time delay can be written
purely as a function of the image positions.

Finally, there is the possibility raised by \citet{courbin02} that
there is a second lens galaxy within the Einstein ring, which would
obviously invalidate our single-galaxy model.  We discuss this
possibility further in \S~\ref{sec:summary}.

\subsection{Comparison to previous models}
\label{subsec:previous}

\begin{figure}
\figurenum{4}
\plotone{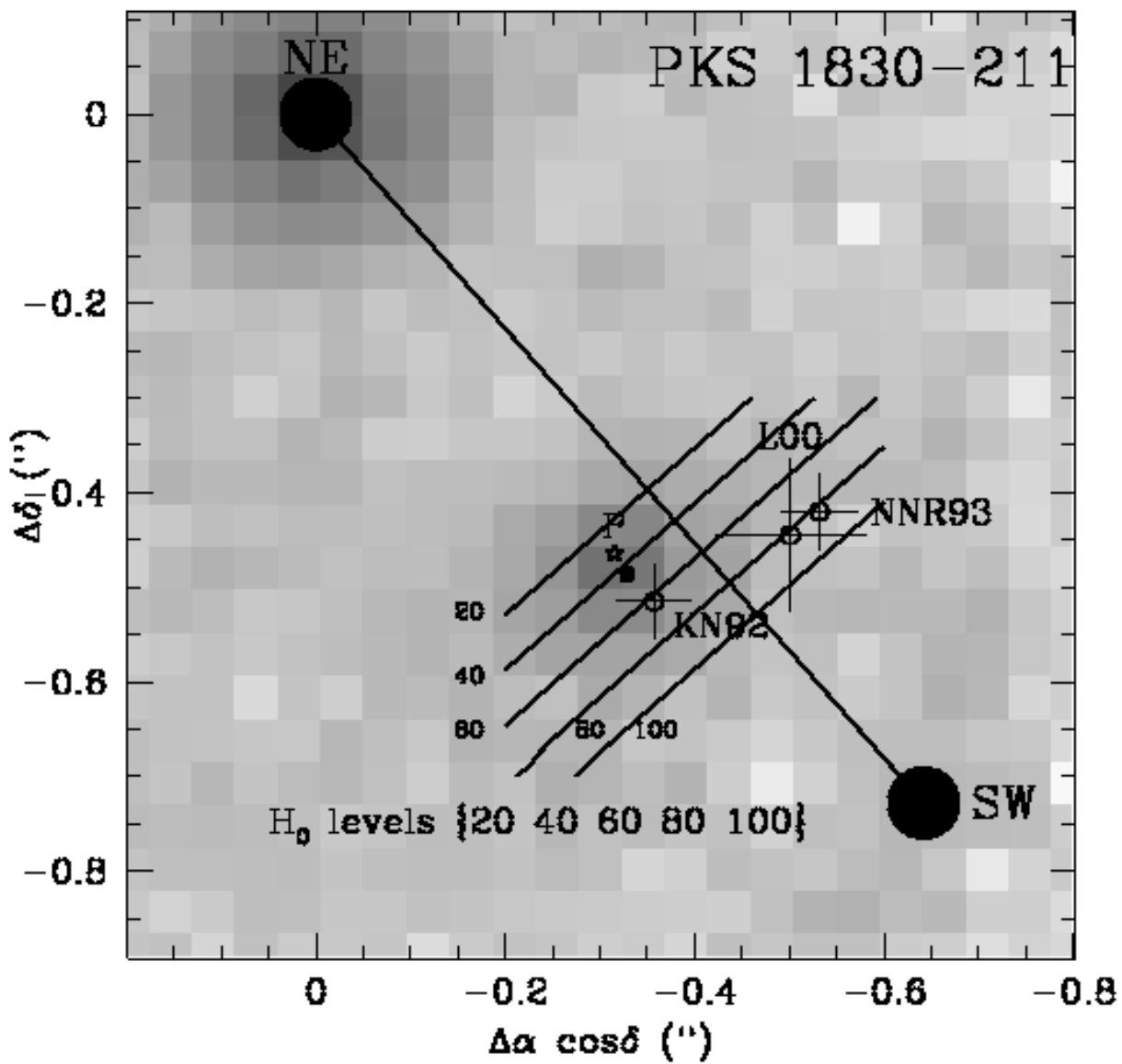}
\caption{ Variation of inferred $H_0$ with lens galaxy position in the
SIE model.  This is an updated version of Figure 5b by
\citet{lehar00}.  Plotted are the lens galaxy positions measured in
this work (filled square) and by Lehar et al.\ (2000, labeled L00);
and the model positions of Nair, Narasimha, \& Rao (1993, NRR93) and
Kochanek \& Narayan (1992, KN92).  The point P was identified by L00
(in an image with a lower signal-to-noise ratio) as either a
foreground star or the bulge of G, and is here seen to be consistent
with the bulge of G. }
\label{fig:positions}
\end{figure}

There have been several previous attempts to model \lens, most of
which differ substantially in the placement of the lens galaxy.
Figure~\ref{fig:positions} is a chart of the lens positions assumed by
various authors, overlayed upon a gray-scale representation of the
same region from the HST $I$-band image.  The filled square shows the
position measured in this work.  The variation in $H_0$ with lens
position (as implied by the SIE model) is shown with contours.  This
figure is an updated version of Figure 5b by \citet{lehar00}.

\citet{kochanek92} invented an inversion algorithm, LensClean, that
determines both the mass model and source structure of lenses with
extended emission, and applied it to radio maps of \lens, with no
constraints on the lens galaxy position.  The resulting lens galaxy
position has a large uncertainty but agrees with the position we have
derived from the HST $I$-band image.

\citet{nair93} used an oblate spheroidal density distribution and a
complex set of constraints inspired by features of VLA maps at 8~GHz
and 15~GHz.  Following \citet{subrahmanyan90}, they required the lens
galaxy to be located near a faint radio component (``E'') that they
argued is an additional image of the background source or emission
from the center of the lens galaxy.  However, although the existence
of E has been verified in other radio images, it is over $0\farcs2$
away from the center of the lens galaxy as revealed by our HST image.
In particular, the center of the lens galaxy and component E are on
opposite sides of the line between A and B, corresponding to a
qualitative difference in the lens models (see below).

All previous authors who have used \lens~to determine the Hubble
constant chose a lens galaxy position close to the incorrect position
of \citet{nair93}.  After measuring the time delay, \citet{lovell98}
applied the model of \citet{nair93} to determine $h=0.65^{+16}_{-9}$.
\citet{koopmans99} obtained $h=0.75^{+18}_{-10}$ using an isothermal
mass model in which the center of the lens galaxy was fixed at the
value of \citet{nair93}.

\citet{lehar00}, who were the first to measure the lens galaxy
position, obtained $h=0.73\pm 0.35$ using an SIE model.  The large
uncertainty was due to the lens galaxy position, which was poorly
constrained, but in agreement with the position of \citet{nair93}.
The difference between their result and ours can be attributed
entirely to the updated position.  In particular, we have assumed the
pointlike object labeled ``P'' by \citet{lehar00} is actually the
bulge of the lens galaxy.  \citet{lehar00} noted this possibility, but
for their analysis they assumed that P was a foreground star.  In an
independent analysis, \citet{courbin02} argue that there is indeed a
spiral lens galaxy but that P is a foreground star superimposed nearly
on the bulge of the galaxy; this possibility is discussed further in
\S~\ref{sec:summary}.

\citet{lehar00} favored a scenario in which the nearby galaxy G2 is an
$L_*$ galaxy located at the lens redshift $z_l=0.886$.  They noted
that the major axis in their SIE model pointed towards G2, suggesting
G2 could naturally explain the ellipticity of the mass model.  By
representing G and G2 as singular isothermal spheres, they achieved a
satisfactory fit, and the resulting ratio of Einstein ring radii
($b_{G2}/b_{G1} = 0.7\pm 0.4$) was consistent with the nearly equal
$H$-band magnitudes and scale lengths of the galaxies.

The new data provide two reasons to reject this interpretation.
First, after moving the lens galaxy to the correct position, the major
axis in the updated SIE model points $91\arcdeg$ away from
G2---exactly the wrong direction for G2 to explain the ellipticity.
Consequently, when we recalculated the SIS+SIS model, $b_{G2}$
converged to zero.

Second, the $VI$ images show that G2 is a spiral galaxy that is larger
in angular size, higher in surface brightness, and bluer than G, all
of which suggest that G2 is at lower redshift.  By summing within the
apertures described in \S~\ref{sec:lensgalaxy} we find that the mean
surface brightness of G2 is $1.7\pm 0.4$ times higher than that of G.
After correcting for Galactic extinction\footnote{The corrections were
computed using the $E_{B-V}$ value of \citet{schlegel98} and assuming
a $R_V=3.1$ extinction law; the results are $A_V=1.4$, $A_I=0.81$, and
$A_H=0.27$.}, we find for G2 that $V-I=0.94\pm 0.28$, as compared to
$V-I>2$ for G.  The blue $V-I$ color of G2 is typical of spiral
galaxies at either $z<0.3$ or $z>1.7$, according to the
spectrophotometric models computed by \citet{lehar00}.  For further
comparison with those models, we computed the $H$-band magnitude of
G2, by by applying the aperture-photometry procedure described in
\S~\ref{sec:lensgalaxy} to the image of \citet{lehar00}. The result
was $H=19.18\pm 0.07$, giving an extinction-corrected magnitude of
$18.9$. This places G2 in the $z<0.5$ section of the $V-I$/$I-H$ plane
\citep[see Fig.\ 3 of][]{lehar00}.

Finally, we note that G2 is the only known candidate for the source of
the $z_a=0.19$ \ion{H}{1} absorption seen by \citet{lovell96}.
Together these facts lead us to favor a scenario in which G2 is
located at $z_a=0.19$.  The extinction-corrected $I$-band magnitude of
G2 is $19.9\pm 0.1$, corresponding to $\sim 0.07L_*$ for a spiral
galaxy at $z_a=0.19$.  Such a small and relatively nearby galaxy would
be irrelevant to lens models with the present degree of accuracy,
contributing a tidal shear and convergence $\gamma \sim \kappa \sim
0.03$.

One argument against this scenario is that the absorption feature
detected by \citet{lovell96} was seen primarily in front of NE, even
though SW is closer in projection to G2 ($2\farcs 5$ vs.\ $2\farcs0$).
However, given the peculiar and patchy distribution of \ion{H}{1} on
kiloparsec scales seen in some nearby galaxies \citep[see,
e.g.,][]{hibbard01}, we do not view this as a serious complication.

\section{Summary and future prospects}
\label{sec:summary}

The new HST $I$-band image confirms that the lens galaxy of \lens~is a
nearly face-on spiral galaxy.  Assuming that the compact component
near the center of this galaxy is the galaxy bulge, we have accurately
measured the galaxy position relative to the quasar images, thereby
completing the basic data for this system.

Furthermore, we have computed $H_0$ given the current best measurement
of the time delay, assuming that there is only one lens galaxy, and
further assuming that the lens galaxy has a flat rotation curve and a
massive dark halo, as appears to be the case for nearby massive spiral
galaxies.  The resulting $H_0$ is lower than the widely accepted value
obtained by the Key Project.  We have described some possible sources
of bias, but to force agreement with the Key Project seems to require
an unnatural mass model for the lens galaxy.

After this work was done, we learned that \citet{courbin02}
independently analyzed the same HST data as presented in
\S~\ref{sec:lensgalaxy} (along with the $IHK$ data of \cite{lehar00}
and a new Gemini $K$-band image).  The main difference between their
treatment of the $VI$ data and ours is that they deconvolved the
images before interpreting them.  Their interpretation differs from
ours in two important respects:

\begin{enumerate}

\item \citet{courbin02} conclude object P is a foreground star rather
than the bulge of the face-on spiral galaxy.  In support of this
claim, P is unresolved and its position in a color-magnitude diagram
is consistent with other bulge dwarf stars for this field.  Based on
the mean density of field stars that were detected in the $I$-band
image ($\sim 0.8$~arcsec$^{-2}$), there is a 10\% chance for a
randomly placed star to lie within $0\farcs2$ of the center of the
galaxy.  The photometry is also consistent with a spiral galaxy with P
as its bulge (see \S~\ref{sec:tullyfisher}), so both interpretations
are reasonable.

\item \citet{courbin02} conclude there is a second lens galaxy, based
on faint and extended flux between P and the SW quasar that is
detected only in the $H$-band image of \citet{lehar00}.  Its position
is rather uncertain (with a quoted error of 80~mas) but is consistent
with the position of radio component E.  In this scenario, the
$H$-band flux and component E are due to a second deflector that is
radio-loud.  In our scenario, these features (if they are real, and
not associated with the quasar) could be due to star formation in the
western spiral arm of lens galaxy.

\end{enumerate}

In short, our interpretation has the virtue of simplicity.  The
interpretation of \citet{courbin02} is complicated but has the merit
of explaining a few puzzling features of the data: a possible offset
between P and the center of the spiral arms, the diffuse $H$-band
flux, and radio component E.  It is not clear to us, from the present
data, how seriously these features should be taken.

If P is indeed a star, the lens galaxy position we report in this
paper is wrong.  Adopting the position of \citet{courbin02} and the
SIE model described in \S~\ref{subsec:update}, the inferred Hubble
constant would rise to $H_0= 107\pm 30$~km~s$^{-1}$.  If, however,
there is a second galaxy between the quasar images, then it may be
impossible to determine $H_0$ from this system because the galaxies
would be difficult to characterize with current telescopes, and the
mass model would be complex.  In particular it is not possible to
state generally (without detailed modeling) whether a two-deflector
model would alleviate or exacerbate the incompatibility of the Key
Project value of $H_0$ and the time delay measured for this system.

Therefore, for the purpose of determining $H_0$ with this system, the
highest priority should be establishing which of these competing
hypotheses are correct.  It would be difficult to establish the
identity of P via spectroscopy, because P is faint and the field is
crowded.  But, as \citet{courbin02} pointed out, it would be possible
to prove P is a star by measuring its proper motion, which should be
$\sim 4$~mas~yr$^{-1}$ for a bulge star (for which the solar reflex
motion of $\sim 200$~km~s$^{-1}$ is dominant).  A negative result
would not be conclusive.  The reality of the second deflector might be
tested by seeking to detect both deflectors in a single
image---perhaps a considerably deeper $I$-band image with the HST's
new Advanced Camera for Surveys.

Assuming for the moment that our interpretation of the HST images is
correct, we suggest the following steps to reduce the systematic
errors and sharpen the determination of $H_0$ from this system.

First, the uncertainty in the time delay should be reduced from 20\%
to 3\% or lower, so that it makes no significant contribution to the
overall uncertainty.  \citet{lovell98} noted that their full light
curves were consistent with a broad range of time delays, ranging from
12 to 30 days.  Only when the data were restricted to a subset
containing a single ``bump'' did the distribution of possible time
delays become approximately Gaussian.  A time delay based on multiple
features in a full set of light curves would be more secure.

Second, the redshift of galaxy G2 ($I=20.7$) needs to be measured
spectroscopically to test whether it is the source of the $z_a=0.19$
\ion{H}{1} absorption. The plausibility of the association with G2
could also be checked by mapping the \ion{H}{1} absorption with a
higher signal-to-noise ratio than the map of \citet{lovell96}, in
order to search for absorption in front of SW, and to measure any
velocity shift.

Third, and most important, the degeneracies of lens models must be
broken by making use of more observational constraints than the basic
data presented in Table~\ref{tbl:data}.  For example, we did not
investigate realistic models for spiral galaxies that include
contributions from the bulge, disk, and halo \cite[see,
e.g.,][]{keeton98,koopmans98}.  More generally, many authors have
argued that unless a wide class of parameterized models are considered
\cite[e.g.,][]{kochanek91,bernstein99}, or even a non-parametric model
\cite[e.g.,][]{williams00}, the uncertainty in $H_0$ will be
underestimated.  To investigate a broader range of models, more
constraints are obviously required.  For the simple models described
in \S~\ref{sec:models}, we have ignored two possible sources of such
constraints: the milliarcsecond-scale radio morphologies of the quasar
cores, and the Einstein ring.

The quasar cores have been mapped with very long baseline
interferometry (VLBI) at radio frequencies ranging from 843~MHz to
43~GHz.  Maps with angular resolution $\gtorder 10$~mas show the
expected parity-reversed substructure in NE and SW \cite[see, e.g.,][]
{jauncey91,patnaik96}, but maps with higher angular resolution are
difficult to interpret.  Component SW, at least, appears to be
scatter-broadened by plasma in the lens galaxy
\citep{jones96,guirado99}.  It has been claimed that both NE and SW
exhibit intrinsic variability in source structure \citep{jin99}.
Furthermore, NE has a $\sim 10$~mas linear jet with no obvious
counterpart in SW \citep{garrett96,guirado99}, which may indicate the
presence of small-scale substructure in the lens.  It is hard to see
how the existing observations can be dramatically improved with
current telescopes, but a modeling effort may help to determine
whether mass subtructure is required, and if so, whether the mass
scale of the perturbation is large enough to affect the time delay.

The geometry of the Einstein ring is probably the best hope for
breaking the remaining degeneracies of lens models.  The situation is
analogous to the case at optical wavelengths, where the stretched
images of quasar host galaxies have been used to break some
degeneracies in mass models of the time delay lens PG~1115+080
(Kochanek, Keeton, \& McLeod 2001; although see Saha \& Williams
2001).  We have made some preliminary attempts to use the existing VLA
maps for this purpose, by employing the curve-fitting algorithm of
\citet{keeton01a}, but with the angular resolution of the present data
it is difficult to measure the ring with the needed accuracy.

There is room for improvement upon existing radio maps of the ring,
which tend to be short snapshots because the radio cores are so bright
($\sim 3$ Jy).  (In most of the VLBI observations, the ring is
resolved out.)  These snapshots are limited in dynamic range by poor
sampling in Fourier space rather than thermal noise.  Major
improvements would follow from deep radio observations with high
angular resolution ($<0\farcs1$) and high dynamic range ($>10^5$).
Methods for the precise interpretation of high-dynamic-range
interferometric observations of lensed sources are well developed and
have been shown to discriminate between different mass models
\cite[see, e.g.,][]{kochanek92,kochanek95,chen95,ellithorpe96}.

\acknowledgments We thank Aaron Cohen and Jim Lovell for help with
this research. We are grateful to Steven Beckwith for providing
Director's Discretionary time for this project. This work was
partially supported by NASA through grants DD-8804, GO-7495, and
GO-9133 from the Space Telescope Science Institute. J.N.W.\ is
supported by an NSF Astronomy and Astrophysics Postdoctoral
Fellowship.

\begin{deluxetable}{ccc}
\tabletypesize{\scriptsize}
\tablecaption{Basic data for \lens\label{tbl:data}}
\tablewidth{0pt}

\tablehead{
\colhead{Datum} &
\colhead{Value} &
\colhead{Reference}
}

\startdata
R.A.(NE) (J2000)              & $18\fh 33\fm 39\fs 931$ & \citet{subrahmanyan90} \\
Decl.(NE) (J2000)             & $-21\fdg 03\farcm 39\farcs75$ & \citet{subrahmanyan90} \\
R.A.(SW) -- R.A.(NE)          & $-642\pm 1$~mas & \citet{jin99} \\
Decl.(SW) -- Decl.(NE)        & $-728\pm 1$~mas & \citet{jin99} \\
R.A.(G) -- R.A.(NE)           & $-328\pm 4$~mas & this work \\
Decl.(G) -- Decl.(NE)         & $-486\pm 4$~mas & this work \\
NE/SW magnification ratio     & $1.52\pm 0.05$  & \citet{lovell98} \\
Lens redshift ($z_l$)         & $0.886\pm 0.001$& \citet{wiklind98} \\
Source redshift ($z_s$)       & $2.507\pm 0.002$ & \citet{lidman99} \\
Time delay ($t_{\mathrm{SW}}-t_{\mathrm{NE}}$) & $26^{+4}_{-5}$ days  & \citet{lovell98}
\enddata

\tablenotetext{a}{\citet{wiklind98} measured molecular absorption
lines at redshifts 0.88582 in front of SW, and also an absorption line
in front of NE with a rest-frame velocity shift of
$-147$~km~s$^{-1}$.}

\end{deluxetable}

\begin{deluxetable}{ccc}
\tabletypesize{\scriptsize}
\tablecaption{HST/WFPC2 photometry\label{tbl:mags}}
\tablewidth{0pt}

\tablehead{
\colhead{Component} &
\colhead{F814W $\approx I$} &
\colhead{F555W $\approx V$}
}

\startdata
S1 & $19.33\pm 0.04$ & $21.90\pm 0.18$ \\ 
NE & $21.97\pm 0.05$ & $25.8\pm 0.2$ \\
SW & $>24.9$         & $>26.3$ \\
G  & $22.04\pm 0.25$ & $\geq 24.7$ \\
G2 & $20.69\pm 0.13$ & $22.24\pm 0.25$
\enddata

\end{deluxetable}

\begin{deluxetable}{cc}
\tabletypesize{\scriptsize}
\tablecaption{SIE model parameters\label{tbl:model}}
\tablewidth{0pt}

\tablehead{
\colhead{Parameter} &
\colhead{Value}
}

\startdata
Einstein ring radius ($b$)          & $0\farcs491\pm 0.001$\\
Ellipticity ($\epsilon$)            & $0.091\pm 0.009$ \\
P.A.\ of major axis (E of N)        & $86\fdg1\pm 3\fdg1$ \\
$x_{\hbox{source}} - x_{\mathrm{NE}}$ & $-0\farcs264\pm 0\farcs005$ \\
$y_{\hbox{source}} - y_{\mathrm{SW}}$ & $-0\farcs418\pm 0\farcs005$ \\
Magnification of NE                 & $5.9\pm 0.6$
\enddata

\end{deluxetable}


\begin{thebibliography}{}

\bibitem[Bernstein \& Fischer(1999)]{bernstein99} Bernstein, G.\ \&
Fischer, P.\ 1999, \aj, 118, 48

\bibitem[Bond et al.(1994)]{bond94} Bond, J.R., et al.\ 1994, \prl,
72, 13

\bibitem[Bradac et al.(2001)]{bradac01} Bradac, M., et al.\ 2001,
preprint, astro--ph/0112038

\bibitem[Chen, Kochanek, \& Hewitt(1995)]{chen95} Chen, G., Kochanek,
C.S., \& Hewitt, J.N.\ 1995, \apj, 447, 62

\bibitem[Courbin et al.(1998)]{courbin98} Courbin, F., et al.\ 1998,
\apjl, 499, 119

\bibitem[Courbin et al.(2002)]{courbin02} Courbin, F., et al.\ 2002,
submitted (astro--ph/0202026)

\bibitem[Dalal \& Kochanek(2001)]{dalal01} Dalal, N.\ \& Kochanek,
C.S.\ 2001, preprint (astro--ph/0111456)

\bibitem[Djorgovski et al.(1992)]{djorgovski92} Djorgovski, S., et
al.\ 1992, \mnras, 257, 240

\bibitem[Dolphin(2000)]{dolphin00} Dolphin, A.\ 2000, \pasp, 112, 1397

\bibitem[Eisenstein, Tegmark, \& Hu(1998)]{eisenstein98} Eisenstein,
D.J., Tegmark, M., \& Hu, W.\ 1998, \apjl, 504, 57

\bibitem[Ellithorpe, Kochanek, \& Hewitt(1996)]{ellithorpe96}
Ellithorpe, J.D., Kochanek, C.S., \& Hewitt, J.N.\ 1996, \apj, 464,
556

\bibitem[Falco et al.(1999)]{falco99} Falco, E.E., et al.\ 1999, \apj,
523, 617

\bibitem[Freedman et al.(2001)]{freedman01} Freedman, W., et al.\
2001, \apj, 553, 47

\bibitem[Fruchter \& Hook(2002)]{fruchter02} Fruchter, A.\ \& Hook,
R.N.\ 2002, \pasp, in press (astro--ph/9808087)

\bibitem[Frye, Welch, \& Broadhurst(1997)]{frye97} Frye, B., Welch,
W.J., \& Broadhurst, T.\ 1997, \apj, 478, 25

\bibitem[Garrett et al.(1996)]{garrett96} Garrett, M.A., et al.\ 1996,
in Proc.\ IAU Symp.\ 173, Astrophysical Applications of Gravitational
Lensing, ed.\ C.S.\ Kochanek \& J.N.\ Hewitt (Dordrecht: Kluwer), p.\
189

\bibitem[Gerin et al.(1997)]{gerin97} Gerin, M., et al.\ 1997, \apj,
488, 31

\bibitem[Gorenstein, Shapiro, \& Falco(1988)]{gorenstein88}
Gorenstein, M.V., Shapiro, I.I., \& Falco, E.E.\ 1988, 327, 693

\bibitem[Guirado et al.(1999)]{guirado99} Guirado, J.C., et al.\ 1999,
\aap, 346, 392

\bibitem[Hibbard et al.(2001)]{hibbard01} Hibbard, J.E., et al.\ 2001,
to appear in ASP Conf.\ Series Vol.\ 240, Gas and Galaxy Evolution,
eds.\ J.E.\ Hibbard, M.P.\ Rupen, \& J.H.\ van Gorkom (San Francisco:
ASP), 659 (astro-ph/0110667)

\bibitem[Jauncey et al.(1991)]{jauncey91} Jauncey, D.L., et al.\ 1991,
Nature, 352, 132

\bibitem[Jin et al.(1999)]{jin99} Jin, C., et al.\ 1999, New Astronomy
Reviews, 43, 767

\bibitem[Jones et al.(1996)]{jones96} Jones, D.L., et al.\ 1996,
\apjl, 470, 23

\bibitem[Keeton \& Kochanek(1998)]{keeton98} Keeton, C.R.\ \&
Kochanek, C.S.\ 1998, \apj, 495, 157

\bibitem[Keeton, Kochanek, \& Falco(1998)]{keeton_kochanek98} Keeton,
C.R., Kochanek, C.S., \& Falco, E.E.\ 1998, \apj, 509, 561

\bibitem[Keeton(2001a)]{keeton01a} Keeton, C.R.\ 2001a, preprint,
astro--ph/0102340

\bibitem[Keeton(2001b)]{keeton01b} Keeton, C.R.\ 2001b, preprint,
astro--ph/0111595

\bibitem[Kochanek(1991)]{kochanek91} Kochanek, C.S.\ 1991, \apj, 373,
354

\bibitem[Kochanek(1995)]{kochanek95} Kochanek, C.S.\ 1995, \apj, 445,
559

\bibitem[Kochanek, Keeton, \& McLeod(2001)]{kochanek01} Kochanek,
C.S., Keeton, C.R., \& McLeod, B.A.\ 2001, \apj, 547, 50

\bibitem[Kochanek \& Narayan(1992)]{kochanek92} Kochanek, C.S.\ \&
Narayan, R.\ 1992, \apj, 401, 461

\bibitem[Koopmans et al.(1998)]{koopmans98} Koopmans, L.V.E., et al.\
1998, \mnras, 295, 534

\bibitem[Koopmans \& Fassnacht(1999)]{koopmans99} Koopmans, L.V.E.\ \&
Fassnacht, C.D.\ 1999, \apj, 527, 513

\bibitem[Krist \& Hook(1997)]{krist97} Krist, J.E.\ \& Hook,
R.N. 1997, The Tiny Tim User's Guide, version 4.4 (Baltimore: STScI)

\bibitem[Leh\'{a}r et al.(2000)]{lehar00} Leh\'{a}r, J., et al.\ 2000,
\apj, 536, 584

\bibitem[Lidman et al.(1999)]{lidman99} Lidman, C., et al.\ 1999,
\apjl, 514, 57

\bibitem[Lovell et al.(1996)]{lovell96} Lovell, J.E.J., et al.\ 1996,
\apjl, 472, 5

\bibitem[Lovell et al.(1998)]{lovell98} Lovell, J.E.J., et al.\ 1998,
\apjl, 508, 51

\bibitem[Metcalf \& Zhao(2001)]{metcalf01} Metcalf, R.\ \& Zhao, H.\
2001, preprint, astro--ph/0111427

\bibitem[Nair, Narasimha, \& Rao(1993)]{nair93} Nair, S., Narasimha,
D., \& Rao, A.P.\ 1999, \apj, 407, 46

\bibitem[Patnaik \& Porcas(1996)]{patnaik96} Patnaik, A.R.\ \& Porcas,
R.W.\ 1996, in Proc.\ IAU Symp.\ 173, Astrophysical Applications of
Gravitational Lensing, ed.\ C.S.\ Kochanek \& J.N.\ Hewitt (Dordrecht:
Kluwer), p.\ 305

\bibitem[Rao \& Subrahmanyan(1988)]{rao88} Rao, A.\ \& Subrahmanyan,
R.\ 1988, \mnras, 231, 229

\bibitem[Refsdal(1964)]{refsdal64} Refsdal, S.\ 1964, \mnras, 128, 307

\bibitem[Riess(2000)]{riess00} Riess, A.\ 2000, Instrument Science
Report WFPC2 00--04 (Baltimore: STScI)

\bibitem[Rubin, Thonnard, \& Ford(1978)]{rubin78} Rubin, V.C.,
Thonnard, N., \& Ford, W.K.\ Jr.\ 1978, \apjl, 225, 107

\bibitem[Sakai et al.(2000)]{sakai00} Sakai, S., et al.\ 2000, \apj,
529, 698

\bibitem[Saha \& Williams(2001)]{saha01} Saha, P.\ \& Williams, L.\
2001, \aj, 122, 585

\bibitem[Schechter(2000)]{schechter00} Schechter, P.L.\ 2000, to be in
Proc.\ IAU Symp.\ 201, New Cosmological Data and the Values of the
Fundamental Parameters, ed.\ A.N.\ Lasenby \& A.\ Wilkinson (San
Francisco: ASP)

\bibitem[Schlegel, Finkbeiner, \& Davis(1998)]{schlegel98} Schlegel,
D.J., Finkbeiner, D.P., \& Davis, M.\ 1998, \apj, 500, 525

\bibitem[Sofue \& Rubin(2001)]{sofue01} Sofue, Y.\ \& Rubin, V.\ 2001,
\araa, 39, 137

\bibitem[Subrahmanyan et al.(1990)]{subrahmanyan90} Subrahmanyan, R.,
et al.\ 1990, \mnras, 246, 263

\bibitem[Swift, Welch, \& Frye(2001)]{swift01} Swift, J.J., Welch,
W.J., \& Frye, B.L.\ 2001, \apjl, 549, 29

\bibitem[Tully \& Fisher(1977)]{tully77} Tully, R.B.\ \& Fisher, J.R.\
1977, \aap, 54, 661

\bibitem[Vogt et al.(1996)]{vogt96} Vogt, N., et al.\ 1996, \apjl,
465, 15

\bibitem[Vogt et al.(1997)]{vogt97} Vogt, N., et al.\ 1997, \apjl,
479, 121

\bibitem[Wiklind \& Combes(1996)]{wiklind96} Wiklind, T.\ \& Combes,
F.\ 1996, Nature, 379, 139

\bibitem[Wiklind \& Combes(1998)]{wiklind98} Wiklind, T.\ \& Combes,
F.\ 1998, \apj, 500, 129

\bibitem[Williams \& Saha(2000)]{williams00} Williams, L.\ \& Saha,
P.\ 2000, \aj, 119, 439

\bibitem[Witt, Mao, \& Keeton(2000)]{witt00} Witt, H.J., Mao, S., \&
Keeton, C.R.\ 2000, \apj, 544, 98

\bibitem[Wu et al.(1995)]{wu95} Wu, X.-P., et al.\ 1995, \apjl, 448,
65

\bibitem[Ziegler et al.(2001)]{ziegler01} Ziegler, B.L., et al.\ 2001,
preprint (astro-ph/0111146)

\end{thebibliography}
\end{document}